\documentclass[12pt,times]{article}
\usepackage{enumerate}
\usepackage{amsmath}
\usepackage{amssymb}
\usepackage{alltt}
\usepackage{setspace}
\usepackage{subfigure}
\usepackage{sidecap}
\usepackage{sectsty}
\partfont{\large}
\sectionfont{\large}
\subsectionfont{\small}
\usepackage[left=2cm,top=2cm,right=3cm,head=1cm,foot=1cm]{geometry}
\numberwithin{equation}{section}
\let\oldsqrt\sqrt
\def\sqrt{\mathpalette\DHLhksqrt}
\def\DHLhksqrt#1#2{%
\setbox0=\hbox{$#1\oldsqrt{#2\,}$}\dimen0=\ht0
\advance\dimen0-0.2\ht0
\setbox2=\hbox{\vrule height\ht0 depth -\dimen0}%
{\box0\lower0.4pt\box2}}
\newcommand{\al}{\alpha}
\newcommand{\g}{\gamma}
\newcommand{\e}{\varepsilon}
\newcommand{\ta}{\theta}

\newcommand{\G}{\Gamma}

\newcommand{\ph}{\varphi}
\newcommand{\da}{\dagger}
\newcommand{\pa}{\partial}

\newcommand{\la}{\mathcal L}

\newcommand{\M}{\mathcal M}

\newcommand{\x}{\times}

\newcommand{\ml}{\left(\begin{matrix}}
\newcommand{\mr}{\end{matrix}\right)}

\newcommand{\V}{\mathcal V}

\newcommand{\bra}{\langle}
\newcommand{\ket}{\rangle}
\newcommand{\tr}{\text{tr}}
\newcommand{\op}{\mathcal O}
\newcommand{\del}{\delta}
\newcommand{\Del}{\Delta}

\newcommand{\zb}{\mathbb Z}
\newcommand{\ka}{\kappa}

\newcommand{\ep}{\epsilon}

\newcommand{\half}{\tfrac{1}{2}}
\newcommand{\third}{\tfrac{1}{3}}

\newcommand{\s}{\sigma}

\newcommand{\sig}{\Sigma(81)}

\newcommand{\SM}{\text{SM}}

\begin{document}
\title{\Large \textbf{Private Higgs at the LHC}}
\author{Yoni BenTov$^{1}$ \and A. Zee$^{1,2}$}
\date{}
\maketitle
\begin{flushleft}
\textit{$^1$\,Department of Physics, University of California, Santa Barbara CA 93106}\\
\textit{$^2$\,Kavli Institute for Theoretical Physics, University of California, Santa Barbara CA 93106}\\
\end{flushleft}
\begin{abstract}
We study the LHC phenomenology of a general class of ``Private Higgs" (PH) models, in which fermions obtain their masses from their own Higgs doublets with $\op(1)$ Yukawa couplings, and the mass hierarchy is translated into a dynamical chain of vacuum expectation values. This is accomplished by introducing a number of light gauge-singlet scalars, the ``darkons," some of which could play the role of dark matter. These models allow for substantial modifications to the decays of the lightest Higgs boson, for instance through mixing with TeV-scale PH fields and light darkons: the simplest version of the model predicts the ratios of partial widths to satisfy $\G(h \to VV^*)_{\text{PH}}/\G(h \to VV^*)_{\text{SM}} \approx \G(h \to \g\g)_{\text{PH}}/\G(h \to \g\g)_{\text{SM}} \leq 1$ and $\G(h \to b\bar b)_{\text{PH}}/\G(h \to b\bar b)_{\text{SM}} \sim \op(1)$, where the inequalities are saturated only in the absence of Higgs mixing with light darkons. An extension of the model proposed previously for generating nonzero neutrino masses can also contribute substantially to $h \to \g\g$ without violating electroweak precision constraints. If the Higgs coupling to fermions is found to deviate from the Standard Model (SM) expectation, then the PH model may be a viable candidate for extending the SM.
\end{abstract}
\section{The Private Higgs Framework}\label{sec:intro}
The Large Hadron Collider (LHC) is presently running at a center-of-mass energy of 8 TeV, and by the end of this year the mechanism of electroweak symmetry breaking may finally be established. The early hints of a Standard Model (SM)-like Higgs boson with mass $m_h \sim 125$ GeV \cite{atlas, cms} have matured into a 5$\s$ discovery \cite{atlas2,cms2}, but much work remains before it is established whether the Higgs sector is that of the SM or of a more elaborate theory. For instance, many models of fermion masses depend on a rich Higgs sector with many scalar resonances. If these models are correct, then finding a light Higgs may be just an initial glimpse into a rich bosonic sector at the TeV scale.
\\\\
An early analysis of the LHC signals showed that they are consistent with the SM but may favor a somewhat ``photon-philic" and ``vector-phobic" Higgs, with an increased branching fraction into $\g\g$ and a reduced one into electroweak vectors, and an SM-like branching fraction into $b\bar b$ \cite{LHCdata}. A later study also suggested deviations from the SM, again with a larger branching fraction for $h \to \g\g$, but favored a ``vector-philic" and slightly ``gluon-phobic" Higgs boson \cite{interpreting higgs}. The apparent diphoton excess seems to persist in the most recent analyses of the post-discovery data, along with a potentially dramatic suppression of the $h \tau^-\tau^+$ vertex \cite{2HDMX}. Inspired by the Private Higgs (PH) framework \cite{PH, nuPH} we study a general class of multi-Higgs models that can naturally accommodate departures from the branching ratios of the SM, with a separate knob to tune for each vertex of the type $hf\bar f$.\footnote{The framework is also such that the quartic self-couplings in the Higgs potential are generally unimportant, so future measurements of parameters in the Higgs potential that turn out to be smaller than expected from the usual double-well structure of the standard approach to electroweak symmetry breaking may point toward an extension of the SM similar to ours.}
\\\\
In the SM, the charged fermion masses arise from the Higgs doublet with vev $v \sim 246$ GeV that gives mass to the $W^\pm$ and $Z$ gauge bosons. The top mass $m_t \sim 173$ GeV has a Yukawa coupling of $\op(1)$, and the other charged fermion masses are accommodated by fitting the Yukawa couplings to their required empirical values. As a result, the huge ratio $m_t/m_e \sim 10^5$ is obtained from a huge ratio of dimensionless Yukawa couplings, $y_t/y_e \sim 10^5$, without any insight as to why the non-top fermions might be so much lighter than the top quark. 
\\\\
The motivation for proposing that each fermion should obtain mass from its own ``private" Higgs field was precisely to address this issue: the idea is that the observed fermion masses can be accommodated in an extension of the SM with all Yukawa couplings being $\op(1)$. The PH framework is to introduce one Higgs doublet for each SM charged fermion:
\begin{equation}
(\phi_f)_i = \ml \phi_f^+\\\phi_f^0\mr \sim (2,+\half)\text{ under } SU(2)\x U(1)\;,\;\; f = u,c,t; d,s,b; e,\mu,\tau\;.
\end{equation}
These Higgs fields are assumed to have flavor-diagonal Yukawa interactions\footnote{This can be achieved by imposing a suitable discrete symmetry \cite{glashow weinberg, wilczek zee}. For example, focusing on the first generation of quarks for concreteness, we could impose a series of internal parities \cite{PH}:
\begin{align}
&K_d: \phi_d,\bar d,S_d \to -(\phi_d,\bar d,S_d)\qquad K_u: \phi_u,\bar u,S_u \to -(\phi_u,\bar u,S_u) \nonumber\\
& K_1: \phi_d,\phi_u,\ml u\\ d \mr,S_d,S_u \to -(\phi_d,\phi_u,\ml u\\ d \mr,S_d,S_u )
\end{align}
and similarly for the second generation. For the third generation we would necessarily impose $K_b$, but we may leave it up to the reader whether to impose a $K_t$ or a $K_3$, so that the fields $\phi_t,\bar t$, and $(t,b)$ may or may not transform under the flavor symmetry. These parity symmetries are broken once the scalars obtain vevs, and one can perform a phenomenological analysis \cite{PH} to understand what is required to generate a realistic CKM matrix.
} and hence, to leading order, result in a unit CKM matrix\footnote{For example, non-zero CKM angles could arise at one loop \cite{one loop CKM, one loop CKM 2}. A typical contribution to the down-type mass matrix $M_d$ would be of the form $\frac{x}{(4\pi)^2}m_b\frac{\mu^2}{M_>^2}\ln\frac{M_>^2}{M_<^2}$, where $x$ is a product of dimensionless couplings and possibly fermion mass ratios, $\mu^2$ is a combination of vevs or dimension-3 couplings, and $M_{>,<}$ are the masses of the two heaviest particles running in the loop. If the heavy masses do not come predominantly from the Higgs mechanism, then the heavy particles generally contribute negligibly to LHC processes such as gluon-gluon fusion and can be ignored safely.}. In this way, each charged SM fermion $f$ obtains mass only from its associated ``private" Higgs doublet $\phi_f$, so the SM relation $m_f = \tfrac{1}{\sqrt2}y_f^{\text{SM}} v$, with $v \approx 246$ GeV, is replaced by the PH relation
\begin{equation}\label{eq:PH relation}
m_f = \tfrac{1}{\sqrt2}y_f^{\text{PH}}v_f\;,\;\; y_f^{\text{PH}} \sim 1 \implies m_f \sim v_f\;.
\end{equation}
The fermion mass hierarchies are to be obtained by a hierarchy of scales in Higgs vacuum expectation values (vevs) rather than by a hierarchy in dimensionless Yukawa couplings:
\begin{equation}\label{eq:PH yukawa}
y_f^{\text{PH}} = \frac{v}{v_f}\,y_f^{\text{SM}} \sim 1\;.
\end{equation}
These models also include SM-singlet real scalar\footnote{For simplicity we will assume these are all even under parity.} fields:
\begin{equation}\label{eq:SM singlets}
S_x \sim (1,0)\text{ under } SU(2)\x U(1)\;,\;\; x = 1,...,N_S
\end{equation}
one of which may serve as a stable dark matter candidate \cite{nuPH, PH+DM} depending on the particular implementation of the PH framework. The original PH model proposed $N_S = N_f$, the number of charged fermions in the SM. Here we will distinguish between two numbers: the total number of scalars, $N_S$, and the number of ``light" scalars, $n_S$, which will mix substantially with the SM-like Higgs boson. In the spirit of generality we will study the ``minimal" nontrivial\footnote{It is consistent with the PH framework for all SM-singlet scalars to have masses as large as $\sim$ TeV, in which case they may not mix substantially with a light Higgs boson with mass $m_h \sim 125$ GeV. In practice this can always be deduced from our study of the $n_S = 1$ case simply by taking the scalar mass large, or equivalently by setting the appropriate mixing angle to zero.} case $n_S = 1$ explicitly, and make general statements about an unspecified number $n_S > 1$ when appropriate.
\\\\
One usually expects the mass of a Higgs boson to be of the order of its vev, and so a phenomenological challenge for any proposal involving $SU(2)$ doublets with small vevs is to explain why such doublets have not yet been observed. In the PH model, the top Higgs is to be treated as special (by choosing a suitable discrete flavor symmetry), while the mass $M_f$ of each non-top\footnote{The implementation of the PH framework in the original model allows for the relation Eq.~(\ref{eq:PH mass scaling relation}) to hold for $f = t$ as well: if $m^3 \sim v^3 \sim m_t^3$, then we have simply $M_t \sim m_t \sim 10^2$ GeV, just as in the SM. It should be noted that a careful treatment of the relation Eq.~(\ref{eq:PH mass scaling relation}) involves only positive-definite numbers, i.e. $m_f$ and $M_f$ are the physical masses of fermion $f$ and PH field $\phi_f$.} PH doublet $\phi_f$ obeys the following scaling relation:
\begin{equation}\label{eq:PH mass scaling relation}
M_{f}^2 \sim \frac{m^3}{m_f}\qquad(f\neq t)
\end{equation}
The $m^3$ in the numerator of Eq.~(\ref{eq:PH mass scaling relation}) denotes a (flavor-dependent) parameter with dimensions of mass cubed, whose particular value is in general model-dependent but in the PH case is proportional to the vevs of the SM-singlet scalars in Eq.~(\ref{eq:SM singlets}). This is why the model requires various TeV-scale singlets: the relation Eq.~(\ref{eq:PH mass scaling relation}) arises from the interplay between a \textit{positive} mass-squared term for the PH fields and an interaction of the generic form $S_x \phi_t^\da \phi_{f\neq t}$ or $S_x S_y \phi_t^\da \phi_{f\neq t}$, depending on the particular symmetries imposed on the Lagrangian. In particular, one may induce the negative mass-squared instability that breaks $SU(2)\x U(1)$ through the vevs of the $S_x$ fields. 
\\\\
The important relation Eq.~(\ref{eq:PH mass scaling relation}) implies that Higgs doublets with small vevs (associated with light fermions) have large masses, which explains why they have not yet been observed.\footnote{Note that Eqs.~(\ref{eq:PH relation}) and~(\ref{eq:PH mass scaling relation}) predict that the hierarchy in Higgs masses is related by a square root to the hierarchy in fermion masses, so that the severe hierarchy $m_e/m_t \sim 10^{-6}$ is softened to a milder relation $M_{t}/M_{e} \sim 10^{-3}$. } Of course, as is usually the case in models beyond the SM, there are numerous unknown parameters in the model, and we have explored only one particular region of the theoretically available parameter space. The reader is invited to extend the idea of ``privacy" to a different phenomenologically interesting regime.
\\\\
To maintain generality we will not commit ourselves to a particular high-energy realization of the PH paradigm, but instead we will study the generic phenomenological features of these types of models. The simplest implementation of the top and bottom quark sector of the theory is discussed in Section~\ref{sec:top and bottom}, the effective Lagrangian for Higgs physics at the LHC is discussed in Section~\ref{sec:effective lagrangian}, and a conclusion is presented in Section~\ref{sec:end}. For the interested reader, a generalization of the Private Higgs model is discussed in Appendix~\ref{sec:general model}, from which all results in Sections~\ref{sec:intro}-\ref{sec:effective lagrangian} can be derived and further extended to accommodate high-energy completions of the PH framework which may differ from the original proposal.
\section{Top and bottom sector: a new 2HDM+S}\label{sec:top and bottom}
In view of the PH scaling relation Eq.~(\ref{eq:PH mass scaling relation}) let us assume that all Higgs doublets besides the top-Higgs ($\phi_t$) and the bottom-Higgs $(\phi_b)$ are heavy enough to decouple\footnote{Note that, in contrast to the usual multi-Higgs models, the decoupling limit $M^2 \to \infty$ for heavy PH fields remains well-defined within perturbation theory. This is because $M^2 > 0$ is a high-energy parameter that we fix as input, which in view of Eq.~(\ref{eq:PH mass scaling relation}) induces a small vev $v \sim 1/M^2 \ll M$ in the potential at low energy. In contrast, the usual double-well potential would have $M^2 < 0$ be a low-energy (unphysical) parameter, and the induced physical mass would be $m \sim v \sim (-M^2)^{1/2}$. Of course, this argument is somewhat formal since in our model the PH fields must have an upper limit to their mass in order to give the correct SM fermion masses [see Eq.~(\ref{eq:bottom mass})]. The point is simply that the requirement is fixed by data, not by the validity of perturbation theory.} from all TeV-scale processes at the LHC. For simplicity, we will first specialize to the case of only one light SM-singlet scalar: $n_S = 1$. In particular, we assume that the ``bottom" scalar, $S_b \equiv S$, mixes substantially with the light Higgs, while the other scalars are too heavy to be important.\footnote{In the spirit of the original model, we might want to also include a light ``top" scalar, $S_t$, in addition to $S_b$. It is possible to realize the PH framework by taking the mass-squared parameter of $\phi_t$ to be negative, in which case its vev and mass are obtained essentially in the same way as in the SM. Thus we do not necessarily require a light top scalar, $S_t$, and in the interest of generality we first study the minimal case, $n_S = 1$, for which simple exact analytic expressions can be obtained.} The more general case is discussed in Appendix~\ref{sec:general model}.
\\\\
The (relatively heavy) bottom Higgs has a Lagrangian of the form\footnote{In this paper, somewhat contrary to standard notation, the symbol $D_\mu$ always denotes the explicit 2-by-2 matrix
\begin{equation}\label{eq:covariant derivative}
D_\mu = \pa_\mu-ig\mathcal W_\mu\;,\;\; \mathcal W_\mu \equiv W_\mu^a\left( \half \s^a\right)+\frac{g'}{g}B_\mu\left( +\half I\right) = \ml \tfrac{1}{2c}(c^2-s^2)Z_\mu+sA_\mu&\tfrac{1}{\sqrt2}W^+_\mu \\ \tfrac{1}{\sqrt2}W^-_\mu &-\tfrac{1}{2c}Z_\mu \mr\;,
\end{equation}
rather than the more general notation $D_\mu \Phi = [\pa_\mu-ig \sum_{a\,=\,1}^3W_\mu^a(T_R)^a-ig'B_\mu(+kI)]\Phi$ for any field $\Phi$ transforming in the representation $(R,k)$ of $SU(2)\x U(1)$.}:
\begin{equation}
\la_{\text{heavy}} = \phi_b^\da D^\da D\phi_b-\hat M_b^2\phi_b^\da\phi_b-(J_b^\da \phi_b+h.c.)+\op(\phi_b^\da\phi_b)^2
\end{equation}
where we have defined the field-dependent effective mass-squared
\begin{equation}\label{eq:M-hat_b^2}
\hat M_b^2 = M_b^2+\G_{t}\phi_t^\da\phi_t+\half\G_{S}S^2
\end{equation}
and the field-dependent effective source that couples to $\phi_b$:
\begin{equation}\label{eq:J_b}
J_b = (\mu+\g S)S\phi_t+y_b^{\text{PH}} \ml t\\ b \mr\bar b\;.
\end{equation}
We emphasize that, in contrast to the usual negative mass-squared term for a Higgs field, the quantity $M_b^2 > 0$ in Eq.~(\ref{eq:M-hat_b^2}) is (to leading order) the actual squared mass for the physical particles of $\phi_b$. Note that the imposition of a $\zb_2$ parity $S \to -S$ would forbid the bare dimension-3 coupling $\mu$, as in the original proposal \cite{PH, nuPH}, but promoting $S$ to a complex scalar field and changing the $\zb_2$ to a $\zb_3$ could permit a term of the form $S\phi_t\phi_b$ in the Lagrangian \cite{sig81}. For simplicity we will assume that the parameters $\mu$ and $\g$ are real.
\\\\
Let us consider energy scales $E \lesssim M_b$ and perform the path integral over $\phi_b$ in the Gaussian approximation \cite{2 higgs}. Dropping terms of $\op(J^2/M_b^4)$, we find a leading order contribution to the effective Lagrangian for the light fields:
\begin{equation}\label{eq:integrated out phi_b}
\Del \la_{\text{light}}^{\text{eff}} \approx \frac{1}{\hat M_b^2}\left|\, (\mu+\g S)S\phi_t+y_b^{\text{PH}} \ml t\\ b \mr\bar b\;\right|^2\;.
\end{equation}
The main feature of Eq.~(\ref{eq:integrated out phi_b}) is the effective Yukawa interaction for the bottom quark:
\begin{equation}\label{eq:effective bottom yukawa}
\la_b^{\text{eff}} = \frac{(\mu+\g S)S}{\hat M_b^2}y_b^{\text{PH}} (\phi_t^0)^\da b\bar b+h.c.
\end{equation}
When the light fields obtain vevs,\footnote{We fix a gauge in which the pseudoscalar and charged components of $\phi_t$ are set to zero.} 
\begin{equation}\label{eq:shifted phi_t and S}
\phi_t = \ml 0\\ \tfrac{1}{\sqrt2}(v_t+H_t) \mr\;,\;\; S = v_S+\s
\end{equation}
the bottom quark\footnote{The other non-top fermions get mass in the same way, each from its own PH field. By saying that the heavier PH fields ``decouple" from all $\sim$ TeV-scale processes, we mean that their masses are significantly larger than a TeV. However, as shown in Eqs.~(\ref{eq:integrated out phi_b}) and~(\ref{eq:bottom mass}), and more generally in Eqs.~(\ref{eq:effective yukawa}) and~(\ref{eq:light fermion mass}), their $\op(\M_f^{-2})$ effects clearly have observable consequences for low-energy physics. The content of our assumption is that, other than inducing light fermion masses, their effects are subleading to the other effects we discuss in the present section.} gets a mass:
\begin{equation}\label{eq:bottom mass}
m_b = \frac{(\mu+\g v_S)v_S}{\M_b^2} \tfrac{1}{\sqrt2}y_b^{\text{PH}}v_t
\end{equation}
where $\M_b^2 \equiv \bra \hat M_b^2\ket$ [see Eq.~(\ref{eq:M-hat_b^2})]. The above expression for $m_b$ is the explicit realization of Eq.~(\ref{eq:PH relation}) for this class of PH models (and similarly for the other non-top charged fermions, with the replacement $b \to f$). We will refer to the physical $\s$ field as a ``darkon" (irrespective of whether it possesses the appropriate properties to constitute the dark matter of the universe). The presence of a light darkon may suppress interactions with SM particles \cite{hiding higgs, HS mixing} or contribute nontrivially to the invisible width of the Higgs if allowed kinematically \cite{higgs to dark, higgs to dark2, higgs to dark3, higgs to dark4}. 
\\\\
While the bottom quark (and the other light charged fermions) obtains mass as in Eq.~(\ref{eq:bottom mass}), the top quark obtains a mass in the usual way:
\begin{equation}\label{eq:top mass}
\la_{\text{Yuk}} = -y_t \ml t\\ b \mr_i\e^{ij}(\phi_t)_j\bar t+h.c. = -m_t t\bar t+...+h.c.\;,\;\; m_t = \tfrac{1}{\sqrt2}y_tv_t\;.
\end{equation}
As can be seen from Eq.~(\ref{eq:effective bottom yukawa}), the nontrivial dependence on $\phi_t$ and $S$ in the field-dependent parameter $\hat M_b^2$ will induce a deviation from the SM-like Higgs vertex $hb\bar b$. Putting the expansions of Eq.~(\ref{eq:shifted phi_t and S}) into Eq.~(\ref{eq:effective bottom yukawa}), we find, in the absence of mixing between $H_t$ and $\s$, the interactions:
\begin{align}
&\la_{H_t b\bar b} = -\frac{m_b}{v_t}\left(1-\G_t\frac{v_t^2}{\M_b^2} \right) H_t b\bar b+h.c. \label{eq:H_t b b}\\
&\la_{\s b\bar b} = -\Del_b\, \s b\bar b+h.c. \label{eq:sigma b b}
\end{align}
where $\Del_b$ is a dimensionless effective parameter whose particular form will not concern us\footnote{Using Eq.~(\ref{eq:darkon yukawa}) with $n_S = 1$ and $f = b$, we find:
\begin{equation}
\Del_b \approx \frac{m_b}{v_S}\left(1+\frac{\g v_S}{\mu+\g v_S}-\G_S\frac{v_S^2}{\M_b^2} \right)\;.
\end{equation}
}. If $H_t$ and $\s$ do not mix, then $H_t$ serves as the SM-like Higgs boson, and only Eq.~(\ref{eq:H_t b b}) is important. The effective Yukawa coupling in Eq.~(\ref{eq:H_t b b}) exhibits a deviation of $\op(v_t^2/M_b^2)$ from the SM-like $hb\bar b$ vertex. Note that the dimensionless parameter $\G_t$ can have either sign, so that both a larger and smaller $hb\bar b$ vertex can be accommodated.
\\\\
In general, however, the scalar potential may induce nontrivial quadratic mixing between the Higgs field $H_t$ and the singlet field $\s$. The fields are rotated into their mass eigenstate basis by a 2-by-2 orthogonal matrix $\V$:
\begin{equation}\label{eq:2x2 higgs-darkon mixing}
\ml H_t\\ \s \mr = \V\ml h\\ \chi \mr\;,\;\; \V = \ml \cos\ta&-\sin\ta\\ \sin\ta&\cos\ta \mr\;,\;\;\ta = \half \tan^{-1}\!\left( \frac{2\,\del m^2}{m_{H_t}^2-m_{\s}^2}\right)
\end{equation}
where $m_{H_t}^2 \equiv \pa^2 V/\pa v_t^2$, $m_\s^2 \equiv \pa^2 V/\pa v_S^2$, and $\del m^2 \equiv \pa^2 V/\pa v_S\pa v_t$, and $V$ is the effective scalar potential for the light fields, including the additive contribution\footnote{In view of Eq.~(\ref{eq:integrated out phi_b}), the off-diagonal part is at least of order $\sim (\frac{\g v_S}{M_b})^2 v_Sv_t$, which may or may not be negligible depending on the region of parameter space of interest.} from Eq.~(\ref{eq:integrated out phi_b}). In this case we write $H_t = \cos\ta\, h+... $ and $\s = \sin\ta\,h+...$ and find an effective $hb\bar b$ interaction:
\begin{equation}\label{eq:hbb}
\la_{hb\bar b} = -\left\{\frac{m_b}{v_t}\left(1-\G_t\frac{v_t^2}{\M_b^2} \right)\cos\ta+\Del_b\sin\ta \right\}h b\bar b+h.c.
\end{equation}
Moreover, the mixing matrix of Eq.~(\ref{eq:2x2 higgs-darkon mixing}) induces a suppression of the $ht\bar t$ vertex by a factor $\cos\ta$, as can be seen from Eq.~(\ref{eq:top mass}), which affects the production cross section $\s(gg \to h)$.
\\\\
In the spirit of the original model, one might also wish to include the mixing with the ``top" scalar, $S_t = \bra S_t\ket + \s_t$, leading to a 3-by-3 orthogonal matrix $\V$ which in general depends on three mixing angles, like the CKM and PMNS matrices for quarks and neutrinos, respectively. Instead of treating the general case, let us study the simplifying ansatz of a 3-by-3 orthogonal mixing matrix which also happens to be symmetric, and therefore depends on only two independent mixing angles, $\al$ and $\beta$ \cite{sym mixing, hermitian mixing}. In this case, we write
\begin{equation}
\ml H_t\\ \s_t\\ \s_b \mr = \V \ml h\\ \chi\\ \chi' \mr\;,\;\; \V \approx \ml 1-2(\al^2+\beta^2)&2\al&2\beta\\ 2\al&-1+2\al^2&2\al\beta\\ 2\beta&2\al\beta&-1+2\beta^2 \mr+\op(\al^3,\beta^3,\al^2\beta,\al\beta^2)
\end{equation}
where $\al \sim \bra S_t\ket m_b/(m_{H_t}^2-m_{\s_t}^2)$ and $\beta \sim \bra S_b\ket m_b/(m_{H_t}^2-m_{\s_b}^2)$ have been taken somewhat less than 1. For example, if $\bra S_t\ket \sim \bra S_b\ket \sim v \sim 10^2$ GeV with $m_{\s_t}$ and $m_{\s_b}$ being non-degenerate with $m_{H_t}$, then we have $\al \sim \beta \sim m_b/v \sim 10^{-2}$, in which case Higgs-darkon mixing is small. If the darkon masses are close to $m_{H_t}$, then the angles $\al$ and $\beta$ could have non-negligible effects on the Higgs interactions.
\\\\
For the general case of $n_S > 1$ SM-singlet scalars, the mixing matrix $\V$ is promoted to a $(1+n_S)$-by-$(1+n_S)$ orthogonal matrix, and the effective Yukawa interaction of Eq.~(\ref{eq:sigma b b}) should be replaced by a sum over darkon species: $\Del_b\,\s \to \sum_{x\,=\,1}^{n_S}\Del_{b}^x\s_x$. [See Eqs.~(\ref{eq:higgs-darkon mixing}) and~(\ref{eq:darkon f f-bar vertex}).]
\\\\
Having derived the $hb\bar b$ vertex, we are now in a position to discuss the effective Lagrangian for Higgs phenomenology at the LHC.
\section{Effective Lagrangian for Higgs physics}\label{sec:effective lagrangian}
Let $\mu(i)$ denote the inclusive rate for Higgs production times the branching fraction into the final state $i$, normalized by the corresponding quantity in the SM:
\begin{equation}
\mu(i) \equiv \frac{\s(pp\to h)\times B(h \to i)}{\s(pp\to h)_{\text{SM}}\times B(h \to i)_{\text{SM}}}\;.
\end{equation}
The present experimental values for $i = \{\g\g, VV, \tau^-\tau^+, b\bar b\}$ are  \cite{2HDMX}:
\begin{equation}\label{eq:data}
\mu(\g\g) = 2.0_{-0.5}^{+0.4}\;,\;\; \mu(VV) = 1.0_{-0.3}^{+0.3}\;,\;\; \mu(\tau^-\tau^+) = 0.5_{-2.1}^{+1.6}\;,\;\; \mu(b \bar b) = 1.3_{-0.8}^{+1.8}\;.
\end{equation}
To compare our model to data, we need to compute the low-energy effective Lagrangian, the pertinent part of which can be expressed\footnote{The general expression arises from Eqs.~(\ref{eq:effective kinetic})-(\ref{eq:effective potential}).} in the convenient form \cite{interpreting higgs}
\begin{equation}\label{eq:effective higgs lagrangian}
\la_{\text{eff}} = c_W \frac{2m_W^2}{v}hW^+_\mu W^{-\mu}+c_Z\frac{m_Z^2}{v}hZ_\mu Z^\mu -c_G\frac{\al_S}{6\pi v}h\,\tr(G_{\mu\nu}G^{\mu\nu})-c_\g\frac{\al}{\pi v}h\,F_{\mu\nu}F^{\mu\nu}-\sum_{f\neq t}c_f\frac{m_f}{v}h (f\bar f+h.c.)\;.
\end{equation}
In the PH model, the only spin-0 fields charged under $SU(2)$ transform as doublets, just as in the SM, so the $hWW$ and $hZZ$ interactions are modified by a common factor:
\begin{equation}
c_W = c_Z \equiv c_V\;.
\end{equation}
The effective Lagrangian of Eq.~(\ref{eq:effective higgs lagrangian}) is written for energies below $m_t$, for which the top quark has been integrated out. There is also a cubic interaction between the Higgs and the darkon fields, of the form $h\chi\chi$ [see Eq.~(\ref{eq:2x2 higgs-darkon mixing})], which contributes to the invisible width of the Higgs if $m_\chi < \half m_h \sim 63$ GeV.
\\\\
For a light Higgs boson with mass $m_h < 2m_W$, the decay into $b\bar b$ dominates the width into SM particles. From Eq.~(\ref{eq:hbb}) we find the effective coefficient:
\begin{equation}\label{eq:c_b}
c_{b} \approx \left( 1-\G_{t}\frac{v_t^2}{\M_b^2}\right)\V_{th}+\sum_{x\,=\,1}^{n_S}\Del_b^x\V_{xh}\;.
\end{equation}
In Eq.~(\ref{eq:c_b}) we have generalized the situation of Section~\ref{sec:top and bottom} to accommodate $n_S > 1$ light darkons, with a $(1+n_S)$-by-$(1+n_S)$ orthogonal mixing matrix $\V$ that diagonalizes the mass-squared matrix of the CP-even spin-0 states $(H_t, \s_1,...,\s_{n_S})$.
The $hGG$ vertex arises primarily through the 1-loop exchange of a top quark, which is proportional to the $ht\bar t$ vertex.\footnote{As a reference, the SM Higgs branching fractions for $m_h = 125$ GeV are \cite{interpreting higgs, branching fractions}: \begin{align}
&B(h \to b\bar b) = 58\%\;,\;\;B(h \to WW^*) = 22\%\;,\;\;B(h \to \tau^-\tau^+) = 6.3\%\;,\;\; B(h \to c\bar c) = 2.9\%\;, \nonumber\\
&B(h \to ZZ^*) = 2.6\%\;,\;\;B(h \to gg) = 8.6\%\;,\;\; B(h \to \g\g) = 0.23\%\;,\;\; B(h \to Z\g) = 0.15\%\;.
\end{align} 
and the total width is $\G(h \to \text{all}) \approx 4.0$ MeV. Many thorough reviews of Higgs phenomenology can be found in the literature \cite{higgs at LHC, anatomy, h to 2gamma bsm}.} From Eq.~(\ref{eq:top mass}) we see that the only deviation from the SM process arises through the Higgs-darkon mixing matrix $\V$ of Eq.~(\ref{eq:2x2 higgs-darkon mixing}). The coefficient $c_{f\neq b,t}$ for the other light SM charged fermions can be obtained by the replacement $b \to f \neq b,t$ in Eq.~(\ref{eq:c_b}).
\\\\
If the production cross section is not changed substantially [we will return to this point later, see e.g. Eqs.~(\ref{eq:top mass}),~(\ref{eq:2x2 higgs-darkon mixing}), and~(\ref{eq:production})], then the physical quantity which enters in the calculation of the $\mu(i)$ is the branching fraction into mode $i$ normalized by the corresponding SM value:
\begin{equation}\label{eq:rho}
\rho(i) \equiv \frac{B(h \to i)}{B(h \to i)_{\text{SM}}} = \frac{\G_h^{\text{SM}}}{\G_h} \frac{\G(h \to i)}{\G(h \to i)_{\text{SM}}}
\end{equation}
Eq.~(\ref{eq:c_b}) suggests that we might attempt to increase the branching fraction into two photons by suppressing the $h b\bar b$ vertex. The general expression for the Higgs branching fraction into two photons normalized by the SM value is: 
\begin{equation}\label{eq:rho 2photons}
\rho(\g\g) = \left(1-\frac{B_X}{B(h \to \g\g)_{\text{SM}}}\right) + \left(\frac{B(h \to b\bar b)_{\text{SM}}}{B(h \to \g\g)_{\text{SM}}}\right)\eta+\left(\frac{\sum_{i\;\ep\;\text{SM}}'B(h \to i)_{\text{SM}}}{B(h \to \g\g)_{\text{SM}}} \right) \xi
\end{equation}
where we have defined $B(h \to b\bar b) \equiv (1-\eta)B(h \to b\bar b)_{\text{SM}}$ and $\sum_{i\;\ep\;\text{SM}}' B(h \to i) \equiv (1-\xi) \sum_{i\;\ep\;\text{SM}}'B(h\to i)_{\text{SM}}$, where the prime on the sum means $i \neq b\bar b,\g\g$. The quantity $B_X \equiv \sum_{x\,=\,1}^{n_S}B(h \to \chi_x\chi_x)$ measures the branching fraction of the Higgs into light darkons. As a simple illustrative example, consider the case $B_X = 0$ and $\xi = 0$, in which case $\rho(\g\g) \approx 1+252\,\eta$. Thus even a small suppression $\eta \sim 10^{-3}-10^{-2}$ of the branching fraction into $b\bar b$ can increase the signal for $h \to \g\g$ to the value observed by the LHC, which makes sense intuitively because the SM branching fraction for $h \to \g\g$ is so small: $B(h \to \g\g)_{\text{SM}} \sim 10^{-3}$. 
\\\\
Of course, we must examine critically whether $\xi = 0, \eta \neq 0$ is realistic, and whether the production cross section remains unchanged from the SM value. We return to this point in Sections~\ref{sec:spin-1} and~\ref{sec:h+}. An alternative possibility is that $B_X \approx B(h \to \g\g)_{\text{SM}}$, in which case $\rho(\g\g) \approx 252\,\eta + 183\, \xi$ is determined purely by $\eta$ and $\xi$. One might also consider (as a fine-tuned example) the case $B_X \neq 0$ with $\G_h = \G_h^{\text{SM}}$ [i.e., tuning $\sum_{x\,=\,1}^{n_S}\G(h \to \chi_x\chi_x)+\G(h \to b\bar b) = \G(h \to b\bar b)_{\text{SM}}$ with the other partial widths held fixed] shows that $\rho(\g\g) > 1$ can be achieved quite easily if the Higgs has any nonzero invisible width and if $c_b^2 = \G(h \to b\bar b)/\G(h \to b\bar b)_{\text{SM}} \lesssim 1$. However, this case requires that the darkons be light while having essentially no mixing with the SM-like Higgs: this implies that these $n_S$ darkons must be extremely light, with masses $m_\chi \ll m_h \sim 10^2$ GeV.
\\\\
The apparent suppression in $B(h \to \tau^-\tau^+)$ can be explained straightforwardly by the analog of Eq.~(\ref{eq:c_b}) with the replacement $b \to \tau$. With a relatively light $\phi_\tau$ of mass $\M_\tau \sim 2v_t \sim 500$ GeV and a large positive coupling $\G_{t\tau} \sim \sqrt{4\pi}$, we obtain (in the absence of mixing with darkons) $c_\tau \sim 1-\G_{t\tau}(v_t/\M_\tau)^2 \sim10\%$. Of course, given the lack of a theoretical principle that determines the sign of $\G_{t\tau}$, this interaction could also accommodate an enhancement in the rate. As the scale of the PH mass is pushed higher, it becomes increasingly difficult to achieve this suppression (or enhancement) with a perturbative coupling constant and a trivial Higgs-darkon mixing matrix $\V$. The phenomenology due to a vast dark sector is potentially extremely rich and could obscure searches for heavy PH bosons.
\section{Higgs decays into spin-1 particles}\label{sec:spin-1}
In addition to the potential modification to $h \to \g\g$ due to Eqs.~(\ref{eq:rho}) and~(\ref{eq:rho 2photons}), one should study the effective coefficient $c_\g$ as defined by Eq.~(\ref{eq:effective higgs lagrangian}). The $h\g\g$ vertex arises through the competition between the 1-loop diagrams with the $W$ boson running in the loop and the 1-loop diagram with the top quark running in the loop (the top loop is also the dominant contribution to the amplitude for gluon-gluon fusion) \cite{higgs at LHC}. It is of critical importance to note that the two contributions enter with opposite signs, and that the $W$ contribution dominates. We will return to this point in Section~\ref{sec:h+}. The vertices involving $W$ bosons arise from the kinetic term $|D\phi_t|^2$. The interactions contained in this term are modified relative to the SM only by the mixing of the Higgs with the darkons, or in other words only by a factor of the mixing matrix $\V$. Thus the leading deviation of the $hVV$ vertex from the SM contains exactly the same factor as the deviation from the $ht\bar t$ vertex.
\\\\
In principle there is also a nontrivial contribution due to the lightest charged Higgs boson, $\phi_b^\pm$. The sign of the coupling $\G_t$ in Eq.~(\ref{eq:M-hat_b^2}) is not fixed a priori, so it is possible for the charged Higgs contribution in our model to add either to the dominant contribution from the $W$ and thereby enhance the rate, or to the sub-dominant contribution from the top and thereby decrease the rate \cite{higgs underproduction}. However, even for a relatively light bottom-Higgs, this contribution is at most an $\op(1\%)$ effect\footnote{Denoting the contribution from a spin-0 charged particle $\phi$ whose mass arises predominantly from the Higgs mechanism by the standard notation $A_0(\tau_\phi)$ (see, e.g., Eq.~(2.9) in ref.~\cite{h to 2gamma bsm}), we may compute the contribution from our PH field $\phi_b^\pm$ by the replacement $A_0(\tau_\phi) \to [\tfrac{1}{\sqrt2}\G_t v^2/(2M_b^2)]A_0(\tau_\phi) \sim 10^{-1}\G_t/(M_b/v)^2$, where we have used $A_0(0) = \third$. Thus, comparing to the $W$ contribution $A_1(\tau_W) \sim -10$, we find $|A_0(\tau_\phi)/A_1(\tau_W)| \sim 10^{-2}\G_t/(M_b/v)^2$. So even for a relatively light PH, e.g. $M_b \sim 3v$, and a large coupling $\G_t \sim 4\pi \sim 10$, the charged Higgs contribution is only $\sim 1\%$ of the $W$ contribution.} and is therefore negligible.
\\\\
Thus in our model the effective coefficients modifying $hWW, hGG,$ and $h\g\g$ are related to those of the SM by a common factor:
\begin{align}\label{eq:effective couplings}
\frac{c_G}{c_G^{\text{SM}}} \approx \frac{c_V}{c_V^{\text{SM}}} \approx \frac{c_\g}{c_\g^{\text{SM}}} \approx \V_{th}\;.
\end{align}
For the case $n_S = 1$, this is simply the factor $\cos\ta$ from Eq.~(\ref{eq:2x2 higgs-darkon mixing}). In general, provided the darkon masses are close enough to the Higgs mass, the element $\V_{th}$ can be decreased significantly from unity, which would suppress Higgs production from gluon-gluon fusion as well as the partial widths $\G(h \to VV)$ and $\G(h \to \g\g)$. The value $\mu(VV) \approx 1$ is compatible with Eq.~(\ref{eq:effective couplings}) and constrains the Higgs-darkon mixing angles in $\V_{th}$ as a function of the total width $\G_h$ of the Higgs.
\\\\
At this point we are invited to examine critically the possibility of increasing $B(h \to \g\g)$ by suppressing $B(h \to b\bar b)$, i.e. Eq.~(\ref{eq:rho 2photons}). Firstly, we see that since the production cross section is dominated by gluon-gluon fusion, Eq.~(\ref{eq:effective couplings}) tells us that the production cross section is in general suppressed: 
\begin{equation}\label{eq:production}
\frac{\s(pp \to h)}{\s(pp\to h)_{\SM}} \approx \frac{\s(GG \to h)}{\s(GG \to h)_{\SM}} \approx \frac{c_G^2}{(c_G^\SM)^2} \approx \V_{th}^2.
\end{equation} 
This tells us that the ratios $\mu(i)$ are extremely sensitive to the mixing angles in $\V_{th}$, e.g. $\V_{th} = \cos\ta$ for $n_S = 1$. In particular, for the decay into $V = W,Z$ we find:
\begin{equation}
\mu(VV) \approx \V_{th}^4\,\frac{\G_h^\SM}{\G_h}\;.
\end{equation}
The relation $\mu(VV) \approx 1$ can be satisfied if the width is modified according to $\G_h \approx \V_{th}^4 \G_h^\SM$. In view of Eq.~(\ref{eq:effective couplings}), in the simplest PH model the ratio $\mu(\g\g)$ is equal to $\mu(VV)$, which is not what we want if we take seriously the indication that $\mu(\g\g) \sim 2$ while $\mu(VV) \sim 1$. We will see in Section~\ref{sec:h+} that in the extended PH model we can parametrize the ratio $\mu(\g\g)$ in the form
\begin{equation}
\mu(\g\g) \approx \V_{th}^2(\V_{th}^2+\Del)\,\frac{\G_h^\SM}{\G_h}
\end{equation}
with some parameter $\Del$, which could be $\op(1)$.
\\\\
To gain intuition for the effect of Higgs-darkon mixing on the Higgs interaction with fermions relative to its interactions with electroweak vectors, consider the case for which there are $n_S \geq 1$ light darkons whose mixing angles and parameters are all independent of darkon flavor. Then the orthogonality relation $\V_{th}^2+\sum_{x\,=\,1}^{n_S}\V_{xh}^2 = 1$ implies $\V_{xh} = \sqrt{(1-\V_{th}^2)/n_S}$, and Eq.~(\ref{eq:c_b}) can be rewritten in terms of Eq.~(\ref{eq:effective couplings}):
\begin{equation}\label{eq:c_b/c_V}
\frac{c_b}{c_V} \approx c_b^{(0)}+\Del_b^S\sqrt{n_S\left( \frac{1}{c_V^2}-1\right)} \to c_b^{(0)}+\Del_b^S\frac{\sqrt{n_S}}{c_V}
\end{equation}
where we have defined $c_b^{(0)} \equiv \lim_{\V_{th} \to 1,\V_{xh} \to 0} c_b$ and taken $\Del_b^1 \approx \Del_b^2 \approx ... \approx \Del_b^{n_S} \equiv \Del_b^S$ to be independent of darkon flavor. The arrow in Eq.~(\ref{eq:c_b/c_V}) corresponds to taking $c_V \to 0$ while keeping the number of light darkon species $n_S$ fixed.
\section{Charged scalars and $h \to \g\g$}\label{sec:h+}
It was shown that the Private Higgs framework can be extended to include $SU(3)\x SU(2)$ singlet complex scalar fields $\mathcal H^+$ with unit electric charge, whose interactions with the Higgs doublets break lepton number by two units and thereby generate nonzero Majorana neutrino masses at one loop \cite{nuPH, zeemodel}. Being charged but colorless, these fields contribute to the Higgs decay into two photons without affecting the Higgs production rate via $gg \to h$. The decay rate for $h \to \g\g$ contribution can be written as \cite{h to 2gamma bsm}:
\begin{equation}\label{eq:H+ contribution to gamma gamma}
\G_{\g\g} = K \left| 1+3\left( \frac{2}{3}\right)^2\frac{A_t}{A_W}\left[ 1+\del_{\g\g}\right]+...\right|^2
\end{equation}
where $K = |A_W|^2\al^2m_h^3/(256\pi^3v^2)$, $A_{t,W}$ are the standard functions resulting from the top and $W$, respectively, and
\begin{align}
\del_{\g\g} 
&\approx \frac{3}{32}\V_{th}v^2 \frac{\pa}{\pa v_t}\ln(M_{\mathcal H^+}^2)
\end{align}
is the new contribution due to a single $\mathcal H^+$ field. Here we have taken $m_h^2/(4M_{\mathcal H^+}^2) \ll 1$. The ellipsis stands for contributions due to leptons and to light quarks, which are negligible relative to the contributions from the $W$, the top, and from the new physics we will consider. In our model, $A_t = \V_{th}A_t^\SM$ and $A_W = \V_{th}A_W^\SM$ [see Eq.~(\ref{eq:effective couplings})], and when studying Eq.~(\ref{eq:H+ contribution to gamma gamma}) recall that $A_t^\SM/A_W^\SM < 0$.
\\\\
If the ``top Higgs portal" interaction is given explicitly by $\la = -\ka \phi_t^\da\phi_t \mathcal H^+\mathcal H^-$, with $\ka$ a dimensionless coupling, then $\del_{\g\g} = \frac{3}{32}\ka(vv_t/M_{\mathcal H^+}^2)\V_{th} \sim \frac{3}{32}\ka(vv_t/M_{\mathcal H^+}^2)\V_{th} \sim 10^{-1}\ka (v/M_{\mathcal H^+})^2\V_{th}$. Therefore, one finds that for a negative and relatively large (but perturbative) coupling $\ka \sim -\sqrt{4\pi} \sim -3.5$ and negligible darkon mixing $(\V_{th} \sim 1)$, one can cancel out the contribution of the top with $N_{\mathcal H^+} \sim 10$ charged scalars having masses $M_{\mathcal H^+} \sim 2 v \sim 500$ GeV. In this case, one finds $\G_{\g\g}/\G_{\g\g}^{\text{SM}} \sim 1.2$, a value $\sim 20\%$ larger than predicted by the SM but still smaller than the value $\mu(\g\g) \sim 2.0$ indicated by the data. However, this effect coupled with a reduction in the $hb\bar b$ vertex, as discussed previously, can provide an explanation for the diphoton excess should the value $\mu(\g\g) \approx 2$ turn out to be robust. 
\\\\
Alternatively, one is invited to go beyond the original proposal by including $SU(3)\x SU(2)$ singlet scalars with a relatively large electric charge, $|Q| \gg 1$, or by including a relatively large number of $|Q| = 1$ scalars. This latter possibility may arise naturally if the new fields transform as a multiplet under a ``dark" non-abelian gauge group. 
\section{Discussion}\label{sec:end}
We have studied a general class of multi-Higgs models in which each SM fermion obtains its mass $m_f$ only from an associated ``private" Higgs doublet whose vev satisfies $v_f \sim m_f$ and whose mass scales as $M_f \sim m_f^{-1/2}$. Motivated by the discovery of a 125 GeV particle which is presumably an SM-like Higgs boson with potentially modified rates into SM final states, we have integrated out all non-top Higgs fields and studied the effective field theory at low energies. 
\\\\
Perhaps surprisingly, the LHC phenomenology for this class of models can be generally similar to that of the SM. The main distinctions are small deviations from the SM-like $hf\bar f$ vertex $\sim m_f/v$, and a possible overall suppression of Higgs decays due to the presence of light SM-singlet ``darkons" which mix with the Higgs. In relation to the SM, this class of models favors:
\begin{itemize}
\item $\G(h \to VV^*)_{\text{PH}} \leq \G(h \to VV^*)_{\text{SM}}$, where the inequality is saturated only if the Higgs does not mix with any light, CP-even SM-singlet scalar [see Eqs.~(\ref{eq:2x2 higgs-darkon mixing}) and~(\ref{eq:effective couplings})]. This is readily compatible with the value $\mu(VV) \approx 1$ of Eq.~(\ref{eq:data}).
\item $\G(h \to f\bar f)_{\text{PH}} \sim \G(h \to f\bar f)_{\text{SM}}$ for $f \neq t$, where the rate is generally comparable to the SM but may receive $\op(1)$ corrections due to the rich structure in the Higgs potential [see Eq.~(\ref{eq:c_b})]. In particular, the hints of a strongly suppressed $h\tau^-\tau^+$ vertex can arise from a relatively light PH for the $\tau$. 
\item $\G(h \to \g\g)_{\text{PH}} \leq \G(h \to \g\g)_{\text{SM}}$ in the basic PH model, where the rate is essentially proportional to the partial width $\G(h \to VV^*)$ [see Eqs.~(\ref{eq:top mass}) and~(\ref{eq:effective couplings})]. In a simple extension of the PH model motivated by the prospect of radiative Majorana neutrino masses, the so-called ``Zee bosons" $\mathcal H^+$ may increase $\G(h \to \g\g)$ without affecting the production cross section $\s(gg \to h)$.
\end{itemize} 
Here we quote the partial widths because they follow directly from the effective Lagrangian of Section~\ref{sec:effective lagrangian}. To deduce the implications for the LHC, we have discussed in the main text the branching fractions $B(h \to X) = \G(h\to X)/\G(h\to \text{all})$. A reduction in $\G(h \to b\bar b)$ will reduce the total width $\G_h \equiv \G(h\to \text{all})$ and thereby increase the branching fraction into $VV^*$ and $\g\g$. For example:
\begin{equation}
\frac{B(h \to \g\g)_{\text{PH}}}{B(h \to \g\g)_{\text{SM}}} = \left(\frac{\G_h^{\text{SM}}}{\G_h^{\text{PH}}}\right)\frac{\G(h \to \g\g)_{\text{PH}}}{\G(h \to \g\g)_{\text{SM}}}\;.
\end{equation}
As noted, a decrease in $\G_h$ due to a suppression of the $hb\bar b$ vertex can be compensated by allowing for a nonzero invisible width into darkons. Thus it is possible to have $\G_h^{\text{PH}} \approx \G_h^{\text{SM}}$, in which case a ratio $B(h \to \g\g)_{\text{PH}}/B(h \to \g\g)_{\text{SM}} \neq 1$ would arise purely from the modified vertex in Eq.~(\ref{eq:effective couplings}) and the presence of relatively light charged scalars as in Eq.~(\ref{eq:H+ contribution to gamma gamma}).
\\\\\
The production cross section is also modified by the presence of substantial Higgs-darkon mixing [see Eq.~(\ref{eq:production})]. To compare to LHC data, one computes $\s(pp \to h)\times B(h \to i)$ normalized by its SM value, where $i$ is any final state. In particular, with the inclusion of $SU(3)\x SU(2)$ singlet electrically charged scalars, the PH relation between $i = VV$ and $i = \g\g$ takes the form
\begin{align}\label{eq:VV vs 2gamma}
&\mu(VV) = \frac{\V_{th}^4}{\g_h}\;,\;\; \mu(\g\g) = \frac{\V_{th}^2(\V_{th}^2+\Del)}{\g_h} \implies \frac{\mu(\g\g)}{\mu(VV)} = \frac{\V_{th}^2+\Del}{\V_{th}^2}
\end{align}
where we have defined $\g_h \equiv \G_h^{\text{PH}}/\G_h^\SM$. In principle $\g_h$ can take any value, but to fit the values $\mu(\g\g) \sim 2$, $\mu(VV) \sim 1$, $\mu(b\bar b) \lesssim 1$, and $\mu(\tau^-\tau^+) \ll 1$ we take $\g_h \lesssim 1$. In more detail, if the Higgs decays only into SM final states, then we take $\g_h \ll 1$ by suppressing $G(h \to b\bar b)$, and if we allow a nontrivial invisible width into light darkons, then we increase the value of $\g_h$ up to $\lesssim 1$.
\\\\
We see that the issues of principal importance for distinguishing Private Higgs models from the SM or other multi-Higgs extensions using only sub-TeV observables are a precise measurement of the $hf\bar f$ vertex for $f = b,\tau$ and the interplay between collider measurements of the $hWW$ vertex and dark matter experiments that constrain the presence of light spin-0 SM-singlet scalars. It is also of critical importance to confirm or falsify the diphoton excess, which would have implications for the mechanism of neutrino mass generation in our model, although here we have explored only one possibility for breaking lepton number appropriately.
\\\\
As the energy of the LHC is increased to its original design of $\sqrt{s} = 14$ TeV, it may become necessary to include the lightest ``heavy" Higgs bosons in the model, namely those arising from the bottom and tau sectors, $\phi_b$ and $\phi_\tau$, as fully propagating degrees of freedom. In this case the phenomenology becomes extremely rich, not only due to the presence of heavy CP-even Higgs bosons but also due to the presence of the physical pseudoscalars $A_b, A_\tau$ and charged Higgs fields $\ph_b^\pm,\ph_\tau^\pm$. The discovery or exclusion of new heavy spin-0 degrees of freedom at higher energies will help establish whether the Private Higgs framework plays any role in the generation of fermion masses. 
\\\\
\textit{Acknowledgments:}
\\\\
This research was supported by the NSF under Grant No. PHY07-57035. We thank Rafael Porto for very useful discussions and comments while doing this research.
\appendix
\section{General multi-Higgs model with SM singlet scalars}\label{sec:general model}
This appendix is devoted to a generalization of the Private Higgs models and a derivation of the results of the main text, which should be useful for future work.
\\\\
Let $\phi \sim (2,+\half)$ denote a collection of $SU(2)\x U(1)$ Higgs doublets, and $S \sim (1,0)$ a collection of SM-singlet real\footnote{If need be, the real scalars can be packaged into a collection of $\half n_S$ complex scalars, e.g. for $n_S = 2k$: $\mathcal S_x = \tfrac{1}{\sqrt2}(S_x+iS_{x+k})$. For simplicity we will assume that all $S_x$ are even under parity.} scalars, which we split into ``heavy" and ``light" fields as follows:
\begin{align}\label{eq:notation for heavy and light}
&\qquad\{\phi_A\}_{A\,=\,1}^N\qquad \leftarrow \text{``heavy" with }\,M^2 > 0\;,\nonumber\\
&\{\phi_a\}_{a\,=\,1}^n\;,\;\;\{S_x\}_{x\,=\,1}^{N_S}\qquad\leftarrow \text{``light" with $|$vev$|$} < M\;.
\end{align}
The positive $M^2$ for the heavy Higgs fields implies that we can integrate them out and leave behind a low-energy effective theory \cite{2 higgs} involving only the fields $\phi_a$ and $S_x$. We make no a priori assumption about the signs of the mass-squared terms for the $n$ light Higgs fields or the $N_S$ scalars. In the original model \cite{PH} the choice was made to induce electroweak symmetry breaking purely from mass-squared instabilities for the $S_x$ fields, but in general this is not necessary. Here we will take all SM-singlet scalars to be light enough to mix substantially with the light Higgs bosons, and so in this appendix we take $n_S = N_S$, using the notation introduced below Eq.~(\ref{eq:SM singlets}).
\\\\
We begin with the kinetic, ``mass," and ``source" terms for the heavy Higgs doublets:
\begin{equation}\label{eq:heavy lagrangian}
\la_{\text{heavy}} = \sum_{A\,=\,1}^N\left[ \phi_A^\da D^\da D\phi_A-\hat M_A^2\phi_A^\da\phi_A-(J_A^\da\phi_A+h.c.)\right]
\end{equation}
Here we have defined 
the field-dependent effective mass-squared,
\begin{equation}\label{eq:field-dependent mass-squared}
\hat M_A^2 = M_A^2+\half\! \left(\sum_{a,b\,=\,1}^{n}\G_A^{ab}\phi_a^\da\phi_b+h.c.\right)+\half\!\!\sum_{x,y\,=\,1}^{n_S}\G_A^{xy}S_xS_y
\end{equation}
and the field-dependent $SU(2)$-doublet source,
\begin{equation}\label{eq:SU(2) current}
(J_A)_i = \sum_{a\,=\,1}^n\sum_{x\,=\,1}^{n_S}\left( \mu_{aA}^x+\sum_{y\,=\,1}^{n_S}\g_{aA}^{xy}S_y\right)(\phi_a)_iS_x+(J_A^{\text{Yuk}})_i\;.
\end{equation}
It will be convenient to define the field-dependent cubic coupling
\begin{equation}\label{eq:field-dependent cubic}
\hat\mu_{aA}^x \equiv  \mu_{aA}^x+\sum_{y\,=\,1}^{n_S}\g_{aA}^{xy}S_y\;.
\end{equation}
The most general form for the Yukawa current for the SM fermions is
\begin{equation}\label{eq:general yukawa current}
(J_A^{\text{Yuk}})_i  = (Y_d)_A^{II'}\ml u_I\\ d_I \mr_i\bar d_{I'}+(Y_u)_A^{II'}\,\e_{ij}\!\ml u_I^\da\\ d_I^\da \mr^j\bar u_{I'}^\da+(Y_e)^{II'}_A\ml \nu_I\\ e_I \mr_i \bar e_{I'}
\end{equation}
where $I = 1,2,3$ and $I' = 1,2,3$ label the fermion family.
\\\\
For energy scales $E \ll$ min$(\{M_A^2\}_{A\,=\,1}^N)$, we may safely integrate out the heavy Higgs fields to get an additive contribution to the Lagrangian for the light degrees of freedom\footnote{The inverse derivative is to be understood perturbatively:
\begin{equation}\label{eq:inverse derivative}
(D^\da D-\hat M_A^2)^{-1}f(x) = \frac{-1}{\hat M_A^2}\left[ f(x)+D^\da D\left(\hat M_A^{-2}f(x)\right)+D^\da D\left( \hat M_A^{-2}\,D^\da D\left( \hat M_A^{-2}f(x)\right)\right)+...\right]\;.
\end{equation}}:
\begin{equation}
\Del\la_{\text{light}}^{\text{eff}} = -\sum_{A\,=\,1}^N(J_A^\da)^i\left[ \left(D^\da D-\hat M_A^2 \right)^{-1}\right]_i^{\;\;j}(J_A)_j\;.
\end{equation}
The low-energy Lagrangian
\begin{equation}
\la_{\text{light}}^{\text{eff}} = \la_{\text{light}}+\Del\la_{\text{light}}^{\text{eff}}
\end{equation}
obtains contributions\footnote{The contribution to beyond-SM flavor changing processes come from the effective four-fermion interactions:
\begin{equation}
\Del \la_{\text{flavor}}^{\text{eff}} = +\sum_{A\,=\,1}^N\frac{1}{\hat M_A^2}\left[ \left( J_A^{\text{Yuk}}\right)^\da\right]^i (J_A^{\text{Yuk}})_i\;.
\end{equation}
If only one flavor of fermion couples to each heavy Higgs $\phi_A$, then the interactions of $\Del \la_{\text{flavor}}^{\text{eff}}$ are diagonal in flavor and do not pose any phenomenological problems in the model.} to the kinetic terms, Yukawa interactions, and tree-level potential:
\begin{align}
&\Del\la_{\text{kin}}^{\text{eff}} = \sum_{a,b\,=\,1}^n\sum_{x,y\,=\,1}^{n_S} (\phi_a^\da)^iS_x\sum_{A\,=\,1}^N\frac{(\hat\mu_{aA}^x)^*}{\hat M_A^2}(D^\da D)_i^{\;\;j}\left[ \frac{\hat\mu_{bA}^y}{\hat M_A^2}(\phi_b)_jS_y\right]\;, \label{eq:effective kinetic}\\
&\Del\la_{\text{Yuk}}^{\text{eff}} = \sum_{a\,=\,1}^n\left( \sum_{x\,=\,1}^{n_S}S_x\sum_{A\,=\,1}^N\frac{(\hat\mu_{aA}^x)^*}{\hat M_A^2}(J_A^{\text{Yuk}})_i\right)(\phi_a^\da)^i+h.c.\;, \label{eq:effective yukawa}\\
&\Del V_{\text{light}}^{\text{eff}} = -\sum_{a,b\,=\,1}^n\sum_{x,y\,=\,1}^{n_S}\left( \sum_{A\,=\,1}^N\frac{(\hat\mu_{aA}^x)^*\hat\mu_{bA}^y}{\hat M_A^2}\right)S_xS_y\phi_a^\da\phi_b \;.\label{eq:effective potential}
\end{align}
The standard Lagrangian for the light fields $\{\phi_a\}_{a\,=\,1}^n$ and $\{S_x\}_{x\,=\,1}^{n_S}$ supplemented with Eqs.~(\ref{eq:effective kinetic})-(\ref{eq:effective potential}) and the field-dependent parameters of Eqs.~(\ref{eq:field-dependent mass-squared}) and~(\ref{eq:field-dependent cubic}) all together constitute the TeV-scale physics of the generalized Private Higgs models.
\\\\
In these equations we have dropped all covariant derivatives except for the $\op(D^\da D)$ contribution to the kinetic term, which in principle will induce a wavefunction renormalization for the light fields. In general, this ``$Z$-factor" will be a non-diagonal matrix in the $(n+n_S)$-dimensional flavor space of light neutral bosons. Moreover, the mass-squared matrix for the CP-even sector will be a non-diagonal $(n+n_S)$-by-$(n+n_S)$ real symmetric matrix. Diagonalization will result in the following generalization of Eq.~(\ref{eq:2x2 higgs-darkon mixing}):
\begin{equation}\label{eq:higgs-darkon mixing}
\ml H_1\\ \vdots \\ H_n\\ \s_1\\ \vdots\\ \s_{n_S} \mr = \V\ml h\\ \mathcal H_1\\ \vdots\\ \mathcal H_{n-1}\\ \chi_1\\ \vdots\\ \chi_{n_S} \mr
\end{equation}
where $\V$ is now an orthogonal $(n+n_S)$-by-$(n+n_S)$ matrix. The labeling of the mass eigenstates in Eq.~(\ref{eq:higgs-darkon mixing}) is meant to indicate that the lightest ``mostly Higgs" field is the SM-like Higgs boson $h$, and the heavier ``mostly Higgs" (CP-even) fields are $\mathcal H_1,...,\mathcal H_{n-1}$. The ``mostly singlet" fields are $\chi_1, ..., \chi_{n_S}$, organized by the relative sizes of their masses. However, no assumption is made about the masses of the $h$ and $\mathcal H$ fields relative to the masses of the darkon fields: any number of the darkons could in principle be lighter than $m_h \sim 125$ GeV.
\\\\
In this more general case, the Higgs-darkon decoupling limit corresponds to
\begin{equation}\label{eq:higgs-darkon decoupling}
\V \to \ml V_H&0\\ 0&V_S \mr\qquad(\text{Higgs-darkon decoupling limit})
\end{equation}
where $V_H$ is an $n$-by-$n$ orthogonal matrix and $V_S$ is an $n_S$-by-$n_S$ orthogonal matrix. 
\\\\
The particular example we take in Section~\ref{sec:effective lagrangian} is $A = d,s,b; u,c; e,\mu,\tau$ (and $N = 8$) and $Y_A^{II'} = y_A\,\del_A^I\del^{II'}$ [see Eq.~(\ref{eq:general yukawa current})] for the heavy Higgs fields, and $a = t$ (and $n = 1$) for the light Higgs field (``top-Higgs").
\\\\
The non-top fermions obtain their masses from the effective Yukawa interactions obtained by integrating out the non-top private Higgs fields [see Eq.~(\ref{eq:effective yukawa})]:
\begin{equation}\label{eq:light fermion mass}
m_{f\neq t} = \left| \frac{\tilde\mu_{tf}\!\cdot\!\bra S\ket}{\M_f^2} \;\tfrac{1}{\sqrt2}y_f^{\text{PH}}v_t\right|
\end{equation}
As in the main text [below Eq.(\ref{eq:bottom mass})], we have defined the shifted value of the effective mass-squared $\M_f^2 \equiv \bra \hat M_f^2\ket = M_f^2+\half \G_f^{tt}v_t^2+\half\sum_{x,y}\G_f^{xy}\bra S_x\ket\bra S_y\ket$ and of the cubic coupling $\tilde\mu_{tf}^x \equiv \bra \hat\mu_{tf}^x\ket = \mu_{tf}^x+\sum_{y\,=\,1}^{n_S}\g_{tf}^{xy}\bra S_y\ket$. [See Eqs.~(\ref{eq:field-dependent mass-squared}) and~(\ref{eq:field-dependent cubic}), respectively.]
\\\\
The $hf\bar f$ vertex for the light fermions $f \neq t$ obtains two types of contributions from the effective Yukawa interactions of Eq.~(\ref{eq:effective yukawa}). First there is the term proportional to $H_t$, which reproduces the SM-like interaction plus a small correction due to quartic interactions in the high-energy scalar potential of the form $V \sim \G_{f}^{tt}\phi_t^\da\phi_t \phi_f^\da\phi_f$:
\begin{equation}\label{eq:H_t f f-bar vertex}
\la_{H_t f\bar f} = -\frac{m_f}{v_t}\left(1-\G_f^{tt}\frac{v_t^2}{\M_f^2} \right)H_t\,f\bar f+h.c.
\end{equation}
The second type of contribution comes from the terms in Eq.~(\ref{eq:effective yukawa}) that are linear in darkon fields $\s_x$ before diagonalizing the CP-even scalar mass-squared matrix:
\begin{equation}\label{eq:darkon f f-bar vertex}
\la_{\s f\bar f} = -\sum_{x\,=\,1}^{n_S}\Del_{f}^x\, \s_x f\bar f+h.c.\;
\end{equation}
where the dimensionless effective darkon-fermion Yukawa coupling $\Del_f^x$ is:
\begin{equation}\label{eq:darkon yukawa}
\Del_f^x = \tfrac{1}{\sqrt2}y_f^{\text{PH}}v_t\,\sum_{y\,=\,1}^{n_S}\frac{\tilde\mu_{tf}^y}{\M_f^2}\left\{ \del_{yx}+\bra S_y\ket\left( \frac{\g_{tf}^{yx}}{\tilde\mu_{tf}^y}-\frac{1}{\M_f^2}\sum_{z\,=\,1}^{n_S}\G_f^{zx}\bra S_z\ket\right) \right\}\;.
\end{equation}
Using the rotation matrix in Eq.~(\ref{eq:higgs-darkon mixing}), we deduce from Eqs.~(\ref{eq:H_t f f-bar vertex}) and~(\ref{eq:darkon f f-bar vertex}) the effective $h f\bar f$ coupling:
\begin{equation}\label{eq:c_f}
c_{f\neq t} = \left( 1-\G_{f}^{tt}\frac{v_t^2}{\M_f^2}\right)\V_{th}+\sum_{x\,=\,1}^{n_S}\Del_f^x\V_{xh}\;.
\end{equation}
This is the expression from which the $h b\bar b$ vertex is obtained in Eq.~(\ref{eq:c_b}).
\\\\

\end{document}